\documentclass[twocolumn,superscriptaddress,amsmath,amssymb,aps,prl]{revtex4-2}

\usepackage{graphicx}
\usepackage{dcolumn}
\usepackage{bm}
\usepackage{amsthm}
\newtheorem*{theorem*}{Theorem}

\usepackage[colorlinks,urlcolor=blue,citecolor=blue,linkcolor=blue]{hyperref}

\begin{document}
	\title{Three-dimensional Moir\'e Crystal}

	\author{Ce Wang}
	\affiliation{School of Physics Science and Engineering, Tongji University, Shanghai, 200092, China}
	\author{Chao Gao}
	\email{gaochao@zjnu.edu.cn}
	\affiliation{Department of Physics, Zhejiang Normal University, Jinhua, 321004, China}
 \affiliation{Key Laboratory of Optical Information Detection and Display Technology of Zhejiang, Zhejiang Normal University, Jinhua, 321004, China}
	\author{Jing Zhang}
	\affiliation{State Key Laboratory of Quantum Optics and Quantum Optics Devices, Institute of Opto-Electronics, Collaborative Innovation Center of Extreme Optics, Shanxi University, Taiyuan, P. R. China}
	\affiliation{Hefei National Laboratory, Hefei 230088, China}
	\author{Hui Zhai}
	\affiliation{Institute for Advanced Study, Tsinghua University, Beijing, 100084, China}
	\affiliation{Hefei National Laboratory, Hefei 230088, China}
	\author{Zhe-Yu Shi}
	\email{zyshi@lps.ecnu.edu.cn}
	\affiliation{State Key Laboratory of Precision Spectroscopy, East China Normal University, Shanghai 200062, China}
	\date{\today}
	
	\begin{abstract}
The work intends to extend the moir\'e physics to three dimensions. Three-dimensional moir\'e patterns can be realized in ultracold atomic gases by coupling two spin states in spin-dependent optical lattices with a relative twist, a structure currently unachievable in solid-state materials. We give the commensurate conditions under which the three-dimensional moir\'e pattern features a periodic structure, termed a \textit{three-dimensional moir\'e crystal}. 
We emphasize a key distinction of three-dimensional moir\'e physics: in three dimensions, the twist operation generically does not commute with the rotational symmetry of the original lattice, unlike in two dimensions, where these two always commute. Consequently, the moir\'e crystal can exhibit a crystalline structure that differs from the original underlying lattice. 
We demonstrate that twisting a simple cubic lattice can generate various crystal structures.
This capability of altering crystal structures by twisting offers a broad range of tunability for three-dimensional band structures.
	\end{abstract}

	\maketitle

Large-scale moir\'e patterns emerge when two identical two-dimensional structures are overlaid with an offset. 
During recent years, studies on double-layered two-dimensional materials, such as twisted bilayer graphenes~\cite{cao2018correlated,cao2018unconventional,yankowitz2019tuning,lu2019superconductors, andrei2020graphene,balents2020superconductivity}, twisted transition metal dichalcogenides~\cite{tang2020simulation,regan2020mott} and twisted cuprates~\cite{can2021high,zhao2023time}, reveal that moir\'e patterns can substantially alter the electronic properties of the material. 
Many extraordinary phenomena, including flat bands~\cite{bistritzer2011moire,tarnopolsky2019origin}, unconventional superconductivity~\cite{wu2018theory,isobe2018unconventional,lian2019twisted}, ferromagnetism~\cite{sharpe19science,lin22science}, fractional Chern insulator~\cite{Spanton_2018,Zeng_2023,Morales_2024,Wang_2024,Liu_2024} and fractional quantum anomalous Hall effect~\cite{cai23nature,park23nature,xu2023observation,lu_2024}, have been predicted and discovered through creating moir\'e patterns in two-dimensional materials by twisting. 
The growing interest in twisted bilayer materials has developed a new field in condensed matter physics, known as twistronics~\cite{carr2017twistronics,carr2020electronic}. 


The impact of twistronics goes beyond condensed matter physics. For instance, moir\'e superlattice has been achieved in photonic crystals~\cite{wang2020localization,Lou_2022,Tang_2023,luan2023reconfigurable} and optical lattices~\cite{meng2023atomic}, paving the way for studying moir\'e physics in electromagnetic waves and ultracold atoms. Remarkably, the recent experimental realization of a two-dimensional moir\'{e} lattice in ultracold atomic gases utilizes the two internal states (spins) of atoms instead of two spatially separated layers~\cite{meng2023atomic}. In the experiment, ultracold $^{87}$Rb atoms are loaded in two spin-dependent two-dimensional optical lattices with a relative twisted angle. The wavelengths of these two optical lattices are carefully selected such that each lattice only couples to one of the two spin states of a $^{87}$Rb atom and is transparent to the other spin component. 
Therefore, although atoms with different spins coexist in the same space, their lattice potentials can be viewed as two independent layers, and an external microwave-induced coupling between the two spin states can play the role of inter-layer tunneling. 
In this way, a moir\'e superlattice can be realized in an optical lattice setting, leading to an intriguing new phase observed between the conventional superfluid and Mott insulator transition~\cite{meng2023atomic}. 

\begin{figure*}[t]
  \includegraphics[width=0.95\textwidth]{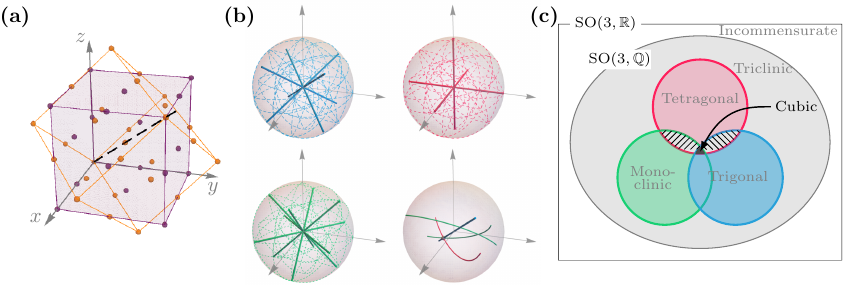}

  \caption{(a) Schematic of an example of a three-dimensional moir\'e crystal. Two sets of cubic optical lattices, represented by the purple and yellow dots, are twisted along the body diagonal $(111)$ direction (dashed line). (b) Three-dimensional twists under which the moir\'e crystal exhibits different rotational symmetry. Each rotation in SO$(3)$ is represented by a point $\mathbf{r}=|\mathbf{r}|\hat{e}_\mathbf{r}$ within a ball of radius $\pi$. Here, $\hat{e}_\mathbf{r}$ represents the twisting axes, and $|\mathbf{r}|$ is the twist angle. The blue, red, and green lines indicate twists where the moir\'e crystals have $C_3$, $C_4$, or $C_2$ symmetry, respectively. The solid lines are positioned along the cubic lattice's symmetry axes. Twists belong to these lines thus preserve the corresponding rotational symmetries. Additionally, twists represented by the dashed curves also lead to rotationally symmetric moir\'{e} crystals. This is due to every twist $R$ on a dashed curve may be transformed into a twist $R’$ on a solid line by $R=R’g$, where $g\in O$. The lower right panel illustrates that the green line can intersect with the blue line, and these three lines can intersect at a single point. Twists represented by these intersection points possess multiple rotational symmetries. (c) The Venn diagram describing the crystal systems of different twists is presented in (b). The red, green, and blue areas represent the Tetragonal, Monoclinic, and Trigonal crystal systems of moir\'e crystal generated by the respective twists in (b). The black area represents the cubic moir\'e crystal generated by the intersection of these lines with different colors. Areas marked by dashed lines are empty sets. The grey area represents a Triclinic crystal system with no rotational symmetry.}  \label{fig1}
\end{figure*}

This letter proposes that this scheme of realizing moir\'e physics in ultracold atoms can be generalized to three dimensions. 
The flexibility in ultracold atom apparatus allows twisting two sets of spin-dependent three-dimensional optical lattices along a generic axis ${\bf L}$ (not any of the principle axes of the original lattice) at a generic angle $\theta$. 
Hence, it can create a \textit{three-dimensional moir\'e crystal} when ${\bf L}$ and $\theta$ satisfy the commensurate condition discussed below. 
This generalization promotes moir\'e physics from two dimensions to three, presenting numerous new possibilities and opening up a new avenue in twistronics.

Here, we highlight a fundamental difference between the moir\'e physics in two and three dimensions. 
The difference stems from the non-abelian nature of the SO$(3)$ rotation group, as opposed to the abelian SO$(2)$ group. 
A two-dimensional twist, i.e., an element in SO$(2)$, always commutes with the rotation symmetry of the original two-dimensional lattice, such as the $C_4$ rotation of a square lattice. 
Consequently, the moir\'e patterns of twisted bilayer square lattices always retain the exact $C_4$ symmetry. 
In contrast, a three-dimensional twist belonging to the non-abelian SO$(3)$ group generally does not commute with the rotation symmetry of the original three-dimensional lattice. 
As a result, a three-dimensional twist can disrupt the rotational symmetry of the original lattice, resulting in a distinct point group symmetry for the three-dimensional moir\'e crystal. 
This property allows a wide range of crystal structures to be generated by twisting the same lattice. It is worth mentioning that there have been previous studies on the moir\'{e} effect on three-dimensional solid-state systems~\cite{20PRR,21PRB,21nano,24PRL}. However, these studies focus on systems of stacked multiple layers of twisted two-dimensional materials. Thus, accessible rotations in these systems are intrinsically two-dimensional and are different from the moir\'{e} crystals considered in this work.

\textit{Physical Model.} A three-dimensional moir\'e crystal can be described by the following Hamiltonian of ultracold atoms with two spin components
\begin{align}
	H=\left(
\begin{matrix}
 \frac{\mathbf{p}^2}{2m}+V_A+\frac{\delta}{2} & \Omega \\
 \Omega & \frac{\mathbf{p}^2}{2m}+V_B-\frac{\delta}{2}
\end{matrix}
\right),\label{Hamiltonian}
	\end{align}
where $m$ is the mass of atoms and $\mathbf{p}$ is the momentum, $\Omega$ represents the coupling between two spin states with a detuning $\delta$, and $V_{A,B}$ are the optical lattice potentials for the two spin states, respectively. 
For illustrative purposes, we consider a three-dimensional cubic lattice, where $V_A({\bf r})$ is written as 
\begin{align}
	V_A(\mathbf{r})=-\frac{V}{4}\left[\cos^2(\pi x)+\cos^2(\pi y)+\cos^2(\pi z)\right].
\end{align}
Here, $V$ stands for the lattice depth, and the recoil momentum has been taken to be $\pi$ for simplicity. 
The optical lattice potential $V_B({\bf r})$ is twisted with respect to $V_A({\bf r})$ by a three-dimensional rotation.
This is characterized by a rotation matrix $R\in\text{SO}(3)$ and a displacement $\mathbf{d}\in\mathbb{R}^3$ as follows:
\begin{align}
V_B(R\mathbf{r}+\mathbf{d})=V_A(\mathbf{r}).\label{VB}
\end{align}
An example of the three-dimensional twisted moir\'e crystals is illustrated in Fig.~\ref{fig1}(a).

\textit{Commensurate Condition.} Similar to the two-dimensional moir\'{e} systems, the moir\'e crystal appears only when the rotation $R$ satisfies certain commensurate conditions, presented by the following theorem.

\begin{theorem*}
The moir\'e pattern formed by overlapping two cubic lattices $V_A({\bf r})$ and $V_B({\bf r})$ forms a three-dimensional periodic lattice structure if and only if $R\in\text{SO}(3,\mathbb{Q})$.
\end{theorem*}

Here, $\text{SO}(3,\mathbb{Q})$ is the set of all three-dimensional special orthogonal matrices with all entries being rational numbers~\cite{SO3R}. In the supplementary material, we show that every rotation matrix $R\in\text{SO}(3,\mathbb{Q})$ can be uniquely parametrized by the axis of rotation $\mathbf{L}\equiv(l_1,l_2,l_3)$ and the rotation angle~\cite{sm}
\begin{align}
	\theta=\arccos\frac{m^2-n^2|\mathbf{L}|^2}{m^2+n^2|\mathbf{L}|^2}.\label{angle}
\end{align}
Here $(l_1,l_2,l_3)$ are three integers with $\text{gcd}(l_1,l_2,l_3)=1$, and $(m,n)$ are two integers with $\text{gcd}(m,n)=1$, where $\text{gcd}$ stands for the greatest common divisor. 
Note that if we take the rotation axis $\mathbf{L}=(0,0,1)$, Eq.~\eqref{angle} becomes $\theta=\arccos\frac{m^2-n^2}{m^2+n^2}$, which coincides the commensurate condition for a twisted bilayer square lattice~\cite{huang2016localization,meng2023atomic}.

\begin{figure}[t]
  \includegraphics[width=\linewidth]{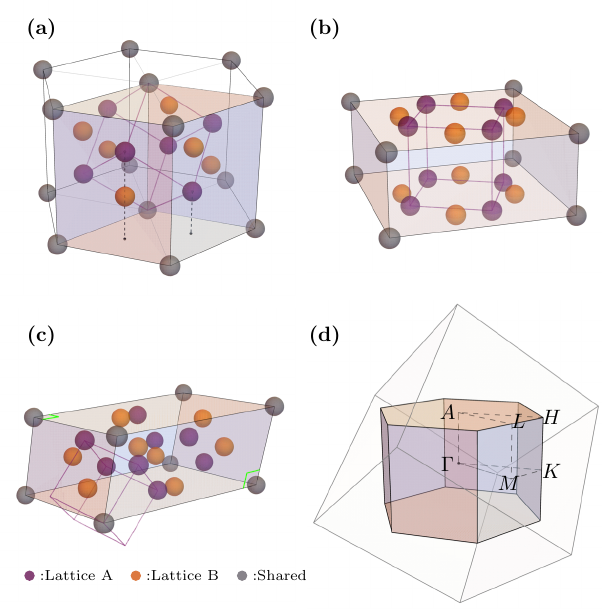}
 
  \caption{(a-c) Lattice structures of twisted cubic lattices with different twisting axes ${\bf L}$ and twisting angle $\theta$. ${\bf L}=(1,1,1)$ and $\theta=\pi/3$ with $(m,n)=(3,1)$ (a); ${\bf L}=(0,0,1)$ and $\theta=\arccos\frac{3}{5}$ with $(m,n)=(2,1)$ (b); ${\bf L}=(1,1,0)$ and $\theta=\arccos\frac{7}{9}$ with $(m,n)=(4,1)$ (c). They belong to the Trigonal, Tetragonal, and Monoclinic crystal systems. The blue, red, and gray dots represent the lattice sites of lattice A, lattice B, and those shared by both lattices. The unit cell is denoted by the shaded prism. (d) The first Brillouin zone (shaded hexagonal prism) and high symmetry points for the reciprocal lattice of the moir\'e crystal are presented in (a). The cube formed by the black edges represents the first Brillouin zone of the original cubic lattice A.}\label{fig2}
\end{figure}

\textit{Proof of the Theorem.} Note that all the lattice vectors of $V_A$ form a three-dimensional Bravais lattice $\mathbb{Z}^3$, i.e. $V_A(\mathbf{r}+\mathbf{a})=V_A(\mathbf{r})$ for all $\mathbf{r}$ if and only if $\mathbf{a}\in\mathbb{Z}^3$. 
Similarly, the lattice vectors of $V_B$ form the set $R\mathbb{Z}^3\equiv\{R\mathbf{a}|\mathbf{a}\in\mathbb{Z}^3\}$. 
One thus concludes that a vector $\mathbf{u}$ is a period of the moir\'e pattern if and only if $\mathbf{u}\in\mathbb{Z}^3\cap R\mathbb{Z}^3$. 
Furthermore, the moir\'e pattern forms a three-dimensional lattice if and only if $\mathbb{Z}^3\cap R\mathbb{Z}^3$ is a three-dimensional Bravais lattice.
 
To prove the theorem, we first show that if the rotation $R\in\text{SO}(3,\mathbb{Q})$, then $\mathbb{Z}^3\cap R\mathbb{Z}^3$ is a three-dimensional Bravais lattice. Consider vectors $R\hat{e}_1$, $R\hat{e}_2$, and $R\hat{e}_3$, where $\hat{e}_1=(1,0,0)^T$, $\hat{e}_2=(0,1,0)^T$, and $\hat{e}_3=(0,0,1)^T$. 
These three vectors are linearly independent and have rational components if $R\in\text{SO}(3,\mathbb{Q})$. 
One can then find integers $n_i$, such that all components of $n_i\cdot R\hat{e}_i$ are integers for each $i=1,2,3$. 
This demonstrates that there exist three linearly independent vectors $n_i\cdot R\hat{e}_i\in\mathbb{Z}^3\cap R\mathbb{Z}^3,\ i=1,2,3$, and hence $\mathbb{Z}^3\cap R\mathbb{Z}^3$ is a three-dimensional Bravais lattice.

To prove the converse, we note that if $\mathbb{Z}^3\cap R\mathbb{Z}^3$ is a three-dimensional Bravais lattice, there must exist three linearly independent integr vectors $\mathbf{u}_1,\mathbf{u}_2,\mathbf{u}_3\in\mathbb{Z}^3\cap R\mathbb{Z}^3$. 
Since $\mathbf{u}_1,\mathbf{u}_2,\mathbf{u}_3\in R\mathbb{Z}^3$, there exist linearly independent integer vectors $\mathbf{v}_1,\mathbf{v}_2,\mathbf{v}_3$ such that $R(\mathbf{v}_1,\mathbf{v}_2,\mathbf{v}_3)=(\mathbf{u}_1,\mathbf{u}_2,\mathbf{u}_3)$. 
By writing $U=(\mathbf{u}_1,\mathbf{u}_2,\mathbf{u}_3)$ and $V=(\mathbf{v}_1,\mathbf{v}_2,\mathbf{v}_3)$, we thus have $R=UV^{-1}\in\text{SO}(3,\mathbb{Q})$. 
This completes the proof.

The above proof shows that the periodicity of a moir\'{e} crystal is determined solely by the twist $R$, i.e. it is not affected by the displacement $\mathbf{d}$. Furthermore, it also provides an algorithm for calculating the unit cells and unit vectors of a three-dimensional moir\'e crystal. 
Note that a vector $\mathbf{u}\in\mathbb{Z}^3$ is a lattice vector if and only if there exists $\mathbf{v}\in\mathbb{Z}^3$ such that $R\mathbf{u}-\mathbf{v}=0$. 
This equation can be viewed as a set of linear Diophantine equations for six integer variables $u_i,v_i,\ i=1,2,3$, where $u_i$ and $v_i$ are the $i$th components of $\mathbf{u}$ and $\mathbf{v}$. 
Given that $R\in\text{SO}(3,\mathbb{Q})$, it can be shown that every solution $(\mathbf{u},\mathbf{v})^T$ can be uniquely expressed as an integer linear combination of three solutions $(\mathbf{u}_1,\mathbf{v}_1)^T$, $(\mathbf{u}_2,\mathbf{v}_2)^T$ and $(\mathbf{u}_3,\mathbf{v}_3)^T$. 
Therefore, $\mathbf{u}_1,\mathbf{u}_2,\mathbf{u}_3$ span the Bravais lattice $\mathbb{Z}^3\cap R\mathbb{Z}^3$ and can be chosen as the unit lattice vectors of the three-dimensional moir\'e crystal~\cite{sm}.

\textit{Symmetry of Moir\'e Crystal.} We now examine the crystal structures of moir\'e crystal by investigating its chiral point group, i.e., the set of all the proper rotational symmetries of the lattice. 
For simplicity, from this point forward, we will focus on the case where the displacement $\mathbf{d}=0$, unless specified otherwise.

It is well known that the rotational symmetry of a cubic lattice $V_A$ is described by the chiral octahedral group $O$. 
This group consists of $24$ rotations~\cite{OctahedralGroup}. 
As $V_B$ is a lattice generated by twisting $V_A$ by a rotation $R$, its chiral point group is then $ROR^{-1}\equiv\{RO_iR^{-1}|O_i\in O\}$. 
Consequently, the chiral point group of the moir\'e crystal can be directly inferred as  $O\cap ROR^{-1}$.

At this point, we can further elaborate on the distinction between the moir\'e lattice in two and three dimensions. 
In two dimensions, because the rotation group $\text{SO}(2,\mathbb{R})$ is abelian, one has $RgR^{-1}\equiv g$ for any $g,R\in\text{SO}(2,\mathbb{R})$. 
Therefore, $O\cap ROR^{-1} \equiv O$ and the moir\'e lattice always maintain the same rotational symmetry as the original lattice. 
In three dimensions, however, one generically has $RgR^{-1}\neq g$ because of the non-abelian nature of SO$(3,\mathbb{R})$. The chiral point group of a three-dimensional moir\'{e} crystal is thus different from that of the original lattice and, more crucially, can be tuned by choosing an appropriate twist $R$. In Fig.~\ref{fig1}(b), we illustrate different classes of twists that lead to different rotation symmetries of the moir\'e crystals. 
We use a three-dimensional ball with radius $\pi$ to represent all SO$(3)$ rotation symmetry, where the direction $\hat{e}_\mathbf{r}$ of a given point denotes its rotation axis and the length $|\mathbf{r}|$ denotes its rotation angle. 

In the first plot, the blue lines represent all the twists that can lead to a $C_3$ symmetric moir\'e crystal belonging to the Trigonal crystal system. Note that the blue solid lines contain all the rotations along the body diagonals such as the $(1,1,1)$ direction. As these body diagonals are the $C_3$ axes of the original cubic lattice $V_A$, any rotation along one of these axes preserves the corresponding rotation symmetry. However, these rotations do not cover all the twists resulting in $C_3$ symmetry. We emphasize that two rotations $R$ and $R'$ should yield the same moir\'{e} crystal if $R=R'g$ for some $g$ belonging to the chiral octahedral group, i.e. the rotation symmetries of $V_A$. This equivalence generates other dashed blue lines in the plot which cover all the rotations leading to moir\'{e} crystals with $C_3$ symmetry. 

\begin{figure}[t]
  \includegraphics[width=\linewidth]{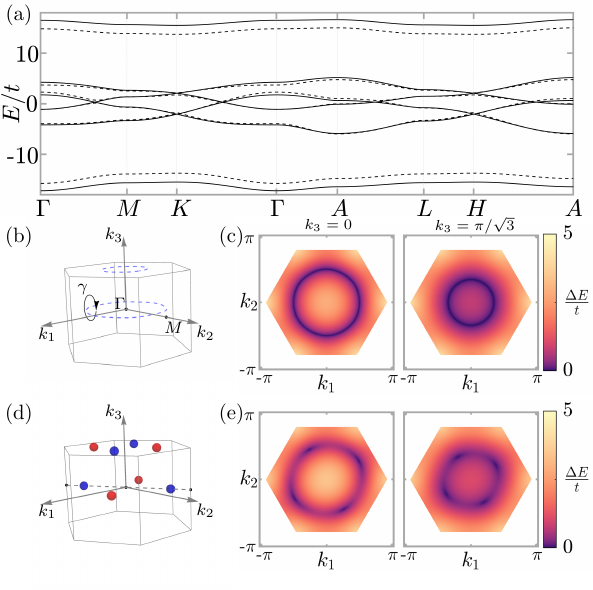}
  \caption{(a) Band structures of the moir\'e crystal with twisting axes ${\bf L}=(1,1,1)$ and twisting angle $\theta=\pi/3$ with $(m,n)=(3,1)$, whose real space structure is shown in Fig. \ref{fig2}(a). The displacements between $V_A$ and $V_B$ are $\mathbf{d}=(0,0,0)$ for solid lines, and $\mathbf{d}=(0,0.1,-0.1)$ for dashed lines. The calculations are based on tight-binding approximation for Hamiltonian~\eqref{Hamiltonian} with parameters $V=6\pi^2/(2m)$, $\Omega=\pi^2/(2m)$ and $\delta=0$. (b,d) Schematic plots showing two topological nodal lines for displacement $\mathbf{d}=(0,0,0)$ (b) and eight Weyl points for displacement $\mathbf{d}=(0,0.1,-0.1)$ (d). The blue/red points indicate the positions for Weyl points with Chern number $+1$/$-1$.  (c,e) Band gaps $E_4(\mathbf{k})-E_3(\mathbf{k})$ for displacement $\mathbf{d}=(0,0,0)$ (c) and $\mathbf{d}=(0,0.1,-0.1)$(e). Left: $k_3=0$ plane; Right: $k_3=\pi/\sqrt{3}$ plane. Note that the dark spots denote the positions of the gapless points.}  \label{fig3}
\end{figure}

Similarly, in the other two plots of Fig.~\ref{fig1}(b), we illustrate the twists that result in a moir\'e crystal with $C_4$ symmetry (red lines) belonging to Tetragonal crystal system, or $C_2$ symmetry (green lines) belonging to Monoclinic crystal system, assuming no other rotational symmetry exists. 
The last plot of Fig.~\ref{fig1}(b) gives examples of how a green line could intersect with a blue line, where the intersection points denote a twist resulting in both $C_3$ and $C_2$ symmetry, also belonging to the Trigonal crystal system. 
Moreover, three lines of different colors can also intersect at the same point, representing a twist that can recover the cubic symmetry. Finally, other twists in SO$(3,\mathbb{Q})$ but does not belong to the lines in Fig.~\ref{fig1}(b) lead to moir\'e crystal with no rotational symmetry, belonging to the Triclinic class. 

The classification of the moir\'e crystal systems is shown in Fig.~\ref{fig1} (c) by the Venn diagram. 
Notably, a rich set of crystal structures, i.e., five out of a total of seven possible crystal systems in three dimensions, emerge through twisting two simple cubic lattices. 
In Fig.~\ref{fig2} we present three examples of different moir\'e crystal belonging to Trigonal, Tetragonal and Monoclinic crystal systems, respectively. The grey points denote the lattice sites shared by A and B lattices, clearly illustrating unit cells with diverse geometries.  

\textit{Band Structures.} The ability to control the crystalline symmetry through three-dimensional twists offers increased flexibility for engineering band dispersion. Here, as a first demonstration of such tunability in three dimensions, we focus on a specific rotation parametrized by twisting along ${\bf L}=(1,1,1)$ by $\theta=\pi/3$. 
The real space moir\'e crystal structure is shown in Fig.~\ref{fig2}(a), and its moir\'e Brillouin zone is depicted in Fig.~\ref{fig2}(d). This is reminiscent of the Brillouin zone of many materials with symmetry-protected nodal lines or nodal points~\cite{15PRL,zhang2017topological,wu2018mgta,Hua_2018,Chan_2019}. Therefore, it inspires us to examine the topological feature in this moir\'e crystal.

In Fig.~\ref{fig3}, we plot the band structures of this moir\'e crystal under a tight-binding approximation~\cite{github}. 
Interestingly, it is discovered that the dispersion can support nontrivial gapless lines between the third and the fourth bands~\cite{sm}, as shown in Fig.~\ref{fig3} (b) and (c). 
These two gapless lines in $k_3=0$ and $k_3=\pi/\sqrt{3}$ planes are topological nodal lines~\cite{burkov2011topological,fang2015topological,yu2015topological,fang2016topological,xu2017topological} protected by either the $\mathcal{PT}$ symmetry or the $\mathcal{M}\sigma_x$ symmetry simultaneously. Here, $\mathcal{P}$ represents the inversion against the origin, 
$\mathcal{T}$ represents the (spinless) time reversal operation, 
$\mathcal{M}$ stands for the reflection against the plane perpendicular to the body diagonal $(111)$ direction, 
and $\sigma_x$ is the spin-flip operation for the two hyperfine spins. 
Both symmetries are maintained when $\mathbf{d}=0$. 
When we introduce a finite perturbation in $\mathbf{d}$, e.g. along the $(0,1,-1)$ direction, both two symmetries are simultaneously broken, which partially gaps the topological nodal lines and results in four pairs of gapless Weyl points~\cite{yan2017topological,armitage2018weyl,sm}, as shown in Fig.~\ref{fig3} (d) and (e). 
These Weyl points all carry nonvanishing Chern numbers with $C=\pm1$, suggesting they are stable against perturbations that preserve the lattice translation symmetries. 

\textit{Conclusion and Outlook.} In summary, we propose to generalize the moir\'e physics to three dimensions. 
We present the commensurate condition for the moir\'e crystal and highlight that different crystal structures of the moir\'e pattern can be realized by twisting the same cubic lattice along different axes and angles. 
The specific features of three-dimensional moir\'e lattices will bring many opportunities for future research. 

Firstly, the complexity of the three-dimensional rotation group allows more flexible band engineering, which can lead to non-trivial band topologies or flat bands. In this work, we have shown the example of generating topological nodal lines and Weyl points through a three-dimensional twist. 
Yet, given the extensive parameter space of the $\text{SO}(3,\mathbb{Q})$ group, more systematic studies of band structures are required for future investigations. 
Secondly, two-dimensional moir\'e systems can exhibit a variety of highly non-trivial correlation effects by introducing interactions~\cite{Kim_2017,balents2020superconductivity,Shimazaki_2020,Wang_2020,Chen_2020,Luo_2021,meng2023atomic}. 
Similar or even more sophisticated correlation effects could occur in three dimensions. 
Thirdly, we have focused on commensurate crystals, while incommensurate twists can also reveal intriguing physics in three dimensions. 
Finally, we have discussed the potential for realizing a three-dimensional moir\'e crystal in ultracold atoms, and a similar realization may also be possible in photonic systems ~\cite{lu2015experimental,zhang2017printing,yang2019realization}.

\textit{Acknowledgement.}
We thank Biao Lian, Zhong Wang, Zengming Meng, Jie Ren, Zhiwei Guo, Yizhou Liu, and Xingze Qiu for inspiring discussion. This work is supported by the National Natural Science Foundation of China~(NSFC) under Grant Nos. 12204352~(CW), 10274342~(CG), 12034011~(JZ), U23A6004~(JZ and HZ), 12004115~(ZYS), the Innovation Program for Quantum Science and Technology under Grant Nos. 2021ZD0302003~(JZ), 2021ZD0302005~(HZ), the National Key R\&D Program of China 2023YFA1406702~(HZ), and the XPLORER Prize~(JZ and HZ), the Natural Science Foundation of Zhejiang Province, China under Grant Nos. LR22A040001~(CG) and  LY21A040004~(CG).
\bibliography{ref}

\begin{thebibliography}{62}
\bibitem [{SO3()}]{SO3R}%
  \BibitemOpen
  \href@noop {} {}\bibinfo {note} {We note that previously we have implicitly assumes that we consider matrices on the field of real numbers. Thus, to avoid confusion, we shall denote $\text{SO}(3)$ by $\text{SO}(3,\mathbb{R})$ hereafter.}\BibitemShut {Stop}%
\bibitem [{Oct()}]{OctahedralGroup}%
  \BibitemOpen
  \href@noop {} {}\bibinfo {note} {They are three $C_4$ axes along the three perpendicular edges (such as the $(100)$ direction) of the cube, four $C_3$ axes along the body diagonals (such as $(111)$), and six $C_2$ axes along the face diagonals (such as $(110)$).}\BibitemShut {Stop}%
\end{thebibliography}

\end{document}